\documentclass[sigconf,nonacm]{acmart}
\AtBeginDocument{%
  }

\usepackage{algorithm}
\usepackage{algpseudocode}
\usepackage{graphicx}
\usepackage{textcomp}
\usepackage[dvipsnames]{xcolor}
\usepackage{booktabs}
\usepackage{soul}
\usepackage{circledsteps}
\usepackage{tabularray}
\usepackage{multirow}
\usepackage{tabularray}
\usepackage{pifont}
\usepackage{wasysym}
\usepackage{hyperref}
\hypersetup{hidelinks}

\soulregister\cite7
\soulregister\autoref7

\begin{document}

\title{R\textsuperscript{3}A: Reliable RTL Repair Framework with Multi-Agent Fault Localization and Stochastic Tree-of-Thoughts Patch Generation}

\newcommand{\rdba}{\emph{R\textsuperscript{3}A}}
\newcommand{\circled}[1]{\raisebox{.5pt}{\textcircled{\raisebox{-.9pt} {#1}}}}

\author{Zizhang Luo}
\email{semiwaker@pku.edu.cn}
\orcid{0000-0002-7276-2317}
\affiliation{%
  \institution{Peking University}
  \city{Beijing}
  \country{China}
}

\author{Fan Cui}
\email{pku_cf@stu.pku.edu.cn}
\orcid{0009-0007-0132-628X}
\affiliation{%
  \institution{Peking University}
  \city{Beijing}
  \country{China}
}

\author{Kexing Zhou}
\email{zhoukexing@pku.edu.cn}
\orcid{0000-0001-7635-3425}
\affiliation{%
  \institution{Peking University}
  \city{Beijing}
  \country{China}
}

\author{Runlin Guo}
\email{guorunlin@buaa.edu.cn}
\affiliation{%
  \institution{Beihang University}
  \city{Beijing}
  \country{China}
}

\author{Mile Xia}
\email{milexia@stu.pku.edu.cn}
\affiliation{%
  \institution{Peking University}
  \city{Beijing}
  \country{China}
}

\author{Hongyuan Hou}
\email{houhy@stu.pku.edu.cn}
\affiliation{%
  \institution{Peking University}
  \city{Beijing}
  \country{China}
}

\author{Yun Liang}
\email{ericlyun@pku.edu.cn}
\orcid{0000-0002-9076-7998}
\affiliation{%
  \institution{Peking University}
  \city{Beijing}
  \country{China}
}

\newcommand{\algorithmautorefname}{Algorithm}

\begin{abstract}

Repairing RTL bugs is crucial for hardware design and verification. Traditional automatic program repair (APR) methods define dedicated search spaces to locate and fix bugs with program synthesis. 
However, they heavily rely on fixed templates and can only deal with limited bugs. As an alternative, Large Language Models with the ability to understand code semantics can be explored for RTL repair. However, they suffer from unreliable outcomes due to inherent randomness and long input contexts of RTL code and waveform.

To address these challenges, we propose R$^3$A, an LLM-based automatic RTL program repair framework upon the basic model to improve reliability. 
R$^3$A proposes the stochastic Tree-Of-Thoughts method to control a patch generation agent to explore a validated solution for the bug. The algorithm samples search states according to a heuristic function to balance between exploration and exploitation for a reliable outcome. Besides, R$^3$A proposes a multi-agent fault localization method to find fault candidates as the starting points for the patch generation agent, further increasing the reliability.
Experiments show R$^3$A can fix 90.6\% of bugs in the RTL-repair dataset within a given time limit, which covers 45\% more bugs than traditional methods and other LLM-based approaches, while achieving an 86.7\% pass@5 rate on average, showing a high reliability.

\end{abstract}

\maketitle

\section{Introduction}

Register‑transfer‑level (RTL) debugging is an indispensable task for hardware design and verification \cite{intel_bug}. In a typical manual debug workflow, engineers first craft testbenches to expose bugs, simulate the design to reproduce them \cite{DBLP:conf/micro/Zhou0LWH23,verilator}, analyze waveforms or assertions to pinpoint the root cause \cite{Zhang_Asgar_Horowitz_2022,DBLP:conf/micro/XuLZLW0024,vasudevan2021learning,5457129}, and finally patch the code. For complex systems, this trial-and-error cycle may require many iterations of simulation and manual inspection, making RTL debugging both time‑consuming and expensive.

Automatic Program Repair (APR)\cite{angelix,Fan_Gao_Mirchev_Roychoudhury_Tan_2023,Xia_Wei_Zhang_2023,swe-agent} has been proven effective in streamlining debugging for both software and hardware. Traditional APR techniques for RTL bug repair combine program synthesis with symbolic execution to search for a predefined solution space\cite{Jiang_Liu_Jou_2005,cirfix,rtl-repair,Ran_Chang_Lin_2003,Chang_Wagner_Bertacco_Markov_2007,strider}. However, these techniques based on fixed templates limit their ability to debug. For example, they can only search for expressions and constants to replace, but can not add or remove registers to deal with timing errors. More importantly, these approaches are unable to comprehend semantics and thus make the searching unguided and inefficient. 

Recently, advances in large language models (LLMs) have highlighted their strong potential for code generation and automated bug fixing \cite{Ahmad_Thakur_Tan_Karri_Pearce_2024b,hdldebugger,fixrag,rtlfixer,uvllm,location,Elnaggar,Fan_Gao_Mirchev_Roychoudhury_Tan_2023,Xia_Wei_Zhang_2023,origen}. LLMs can accept prompts in either natural language or commonly used source code. Crucially, they can capture semantic cues—e.g., recognizing that a signal named reset should only assert during initialization. 
This can potentially enable guided patch generation, which is more efficient than brute-force searching. 
Moreover, because they operate in token space, LLMs can rewrite any portion of the RTL without any manual template constraint.

However, there are two major challenges to \textbf{reliably applying LLM for RTL debugging}, which is the central need for verification.
First, \textbf{LLMs are inherently stochastic}, as they produce tokens by sampling. Improving the base model or fine-tuning\cite{origen,location,hlsdebugger} with an RTL-related dataset can increase the overall quality of the outcome. However, no model can guarantee its correctness, which leaves a large space to gather a better result from randomness. Such approaches are called \textbf{test-time-scaling}\cite{test-time-scaling}, which means improving the model's performance with the cost of increasing inference time. Some works\cite{uvllm, MEIC} horizontally sample the outcome by retrying a few times with the same prompt, automatically verify and score them to get the best patch for the bug. Other approaches\cite{swe-agent, rtlfixer} (LLM agents) sample vertically, which iteratively improves the patch in a continuous dialogue by feeding the response from EDA tools to the LLM, e.g., compiler logs. Vertical sampling simulates the trial-and-error process of a human programmer, but may reach dead ends as they do not explore different paths like horizontal sampling.

Second, \textbf{LLMs require concentrated input}, while RTL codes and waveforms are verbose and sparse in information. Due to their parallel nature, hardware codes are highly unrolled, and waveforms contain a lot of data each cycle. This distracts attention and makes the outcome unstable, e.g., LLMs may guess many error locations throughout the whole code instead of deep thinking about the root cause. Context engineering\cite{context-engineering} comes to help by giving the LLM a focus with a dedicated input. Retrieval-Augmented Generation (RAGs)\cite{hdldebugger, fixrag} can search for fragments in the code and specification that are related to the bug, but it only identifies syntactic similarities instead of logical connections between the fragment and the bug. Tracing the mismatched signal to find suspicious codes\cite{uvllm, strider} also reduces the problem range. However, to find a boundary for tracing, the correctness of the internal signals needs to be predefined, which is impractical, as common verification methods usually only define the correctness of the output signals. 

Addressing these issues, we introduce \rdba, an LLM-driven agent framework for RTL bug repair, which is a test-time-scaling method upon the basic model to improve reliability. \rdba~is composed of three parts: a stochastic tree-of-thought patch generation method, a multi-agent fault localization method, and the Agent-Debugger Interface (ADI) to bridge the gap between text, code and waveform. 
First, inspired by \cite{tree-of-thoughts}, we propose the stochastic tree-of-thought algorithm as the core of the framework, which is an agent workflow controlled by a searcher. The bug fixing trial-and-error process is organized as a search tree to sample the LLM both horizontally and vertically. Unlike basic tree-of-thoughts, which uses DFS or BFS, our stochastic algorithm balances the exploration and exploitation at each iteration by sampling a state to expand based on a heuristic function, \textbf{to provide long-term planning and reliably generate a validated patch for the bug}, overcoming the challenge of stochasticity.

Second, the multi-agent fault localization method gives \textbf{a quick view of the complete code} to find some code fragments as fault candidates. Though they may be inaccurate, the candidates serve as \textbf{good starting points} for the patch generation agent, whose outcome is not restricted to the candidates. 
Finally, the ADI acts as a proxy for LLMs to interact with the debug environment and the EDA tools, reducing the burden on the agents.

Our major contributions are:

\begin{itemize}
    \item \textbf{R\textsuperscript{3}A framework}: We propose a framework to automate RTL program repair with LLM-agents. An Agent-Debugger Interface is designed to adapt the agent to RTL debugging.
    \item \textbf{Patch generation:} We propose an agent workflow with a stochastic tree-of-thoughts sampling method, turning the LLM’s randomness into reliable search and consistently producing valid fixes.
    \item \textbf{Fault localization:} We propose a detection method to process the complete code for good starting points. 
   
\end{itemize}

Evaluated on the RTL-Repair benchmarks \cite{rtl-repair}, experiments show that our approach can fix 90.6\% of bugs in the RTL-repair dataset within a given time limit, which covers 45\% more bugs than traditional methods and other LLM-based approaches, while achieving an 86.7\% pass@5 rate on average, showing a high reliability.

\section{Background and Motivation}

\begin{table}
\caption{Comparison between different APR methods}
\vspace{-0.4cm}
\label{tab:motivation}
\centering
\resizebox{\columnwidth}{!}{
\begin{tblr}{
  colspec={Q[m]Q[m]Q[m]Q[m]},
  cells = {c},
  vline{2} = {-}{},
  hline{2} = {-}{},
  hline{6} = {-}{},
  hline{9} = {-}{},
}
{Name}                        & {Fault\\Localization}     & {Patch\\Generation}                  & {Fix\\Reliability} \\
CirFix\cite{cirfix}             & no                                & synthesis                                       & {limited}                                  \\
RTL-Repair\cite{rtl-repair}     & no                                & solving                                      & {limited}    \\
Strider\cite{strider}           & ~tracing                            & synthesis                                      & {limited}   \\
{\cite{Wu_Zhang_Yang_Meng_He_Mao_Lei_2022}} & {statistics}                    & {no}                                       & {no}                &                              \\
{Location-is-Key\\\cite{location}}  & experience                        & {experience\\translation}                   & {unreliable}                  \\
{RTLFixer\cite{rtlfixer}}         & {analysis\\experience}                       & {experience\\translation\\trial-and-error}                   & {limited}                        \\
{UVLLM\cite{uvllm}}               & {analysis \\tracing \\experience} & {experience\\translation}                   & {limited}                        \\
{R$^3$A\\(this work)}              & {analysis \\experience}          & {experience\\trial-and-error}               & {reliable}     
\end{tblr}
}
\vspace{-0.5cm}
\end{table}

\begin{figure*}[t]
    \centering
    \includegraphics[width=0.8\textwidth]{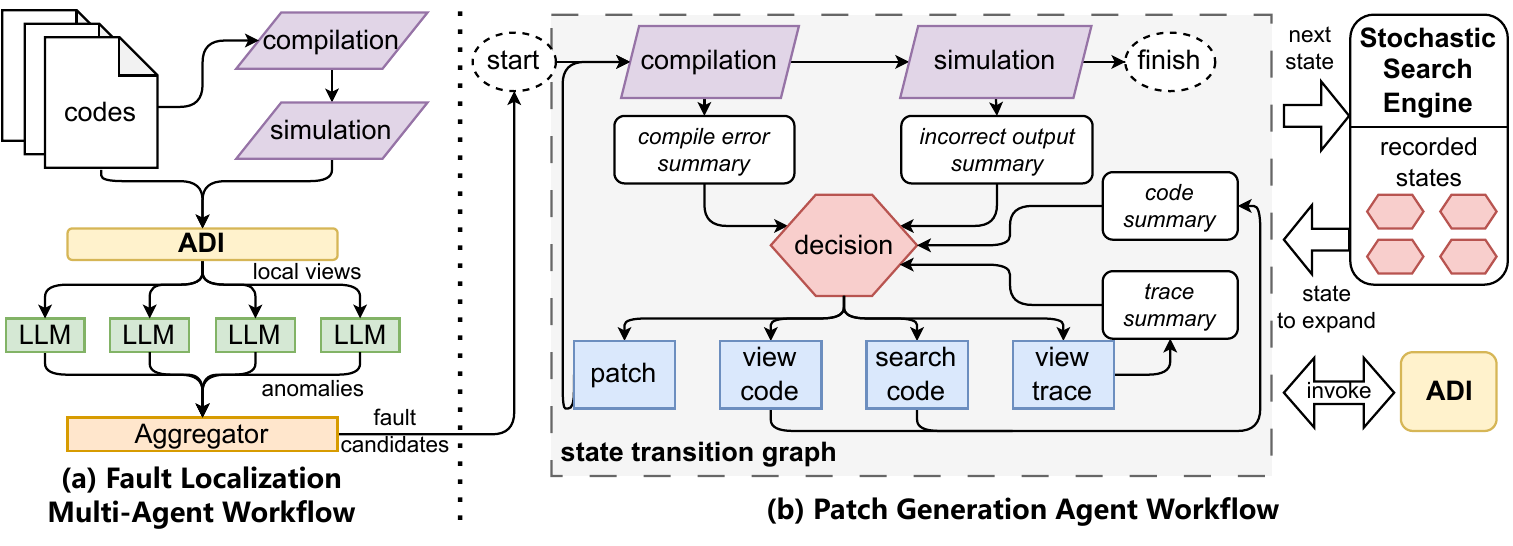}
    \caption{Overview of the \rdba~Framework}
    \label{fig:overview}
    \vspace{-0.5cm}
\end{figure*}

\subsection{RTL Debugging}

RTL debugging aims to restore correct functionality with minimal code modifications and typically proceeds in three stages: bug discovery, fault localization, and patch generation.  
Bug discovery identifies a test case that exposes the fault, either through simulation\cite{DBLP:conf/micro/Zhou0LWH23,verilator} or verification methods\cite{vasudevan2021learning,5457129}. Fault localization then narrows the suspect code region\cite{Ran_Chang_Lin_2003,Jiang_Liu_Jou_2005,Wu_Zhang_Yang_Meng_He_Mao_Lei_2022,location, strider}, and patch generation resolves the fault while minimizing edits\cite{cirfix,rtl-repair,strider,Chang_Wagner_Bertacco_Markov_2007, uvllm}.

This paper targets APR, which assumes bugs are known and focuses on the latter two stages. We define correctness by comparing waveforms during simulation, a method that is both comprehensive for LLMs and standard in RTL debugging. A setting with minimal human guidance is assumed, as in traditional methods: \textbf{no design specification and the LLM observes only the source code and the I/O waveforms of the top module}. 

\subsection{Motivation}

We summarize the existing methods to localize faults and generate patches, and their fix reliability in \autoref{tab:motivation}. Existing traditional methods all have a limited fix range, while LLM-based methods are either impractical or unreliable.

For fault localization, \cite{cirfix} and \cite{rtl-repair} skip this step. \cite{strider} and \cite{uvllm} trace the mismatched signals, which is powerful but not practical as it requires waveforms of the internal signals of a module. \cite{Wu_Zhang_Yang_Meng_He_Mao_Lei_2022} apply spectrum-based approaches from software. LLM-based works can use the learned experience to localize the fault, and some works\cite{rtlfixer, uvllm} use linters\cite{verilator, iverilog} to provide static checks.

Fix reliability is strongly connected with patch generation methods. We define reliability as the probability of finding a valid solution within the given time and resource limit. \cite{cirfix} and \cite{strider} apply program synthesis methods, while \cite{rtl-repair} transforms the problem into an SMT solving problem. However, their fixed capabilities are limited to certain predefined templates. LLM-based methods are naturally unreliable as they rely on hallucinating learn experience. Most of them can translate the specification into code, which is unnatural in large designs to have a specification detailed enough for direct translation. Some works\cite{location} claim to be fine-tuned for better outcomes. Here, we discuss additional methods to further improve their reliability over the base model. \cite{uvllm} samples horizontally by multi-sampling, while \cite{rtlfixer} samples vertically with agent flows to simulate a trial-and-error process by iterative improvement. Their reliability is improved, yet insufficient.

Addressing the challenge of stochasticity and input context quality, a reliable, practical RTL APR agent should have the following features. First, it should make long-term plans to search for the root cause of the bug during the trial-and-error process with learned experience, exploring different paths on a tree while wisely choosing the best for efficiency. Second, it should have some fault candidates as starting points to focus on. The trial-and-error process can address one hypothesis at a time, providing a concise and localized view refined from the broader perspective through experience or analysis, while preserving the ability to find bugs outside of the inaccurate candidates. Third, since RTL debugging requires much more engineering effort than software, the framework should handle RTL code, waveforms, and EDA tools for the agent.

\section{Overview}

\autoref{fig:overview} outlines the major components of \rdba~framework. \autoref{fig:overview}(a) demonstrates the multi-agent workflow for fault localization. First, the codes are compiled and simulated for the error message. The ADI segments the code to form multiple local views. For each local view, an LLM instance is prompted to observe and identify the anomalies. Finally, an aggregator gathers and filters them to produce a set of fault candidates.

As illustrated in \autoref{fig:overview}(b), the patch generation agent runs on a state transition graph, controlled and explored under a stochastic search engine. The fault candidates act as an initial message of the conversation. Then, the agent follows the state transition graph, where each node is a round of conversation. 
Above the agent, the stochastic search engine records all the states that reach the decision node. 
In each iteration, one state is chosen and expanded, running in the state transition graph until it reaches the decision node again.

All actions the agent takes are done through invoking the ADI which provides a text-based interface to interact with the agent.
The ADI gathers problem-centric context for the agents. Combining RAG and traditional methods, the ADI provides a text view for code and waveform fragments centered at a certain line number or cycle count, or related to short search prompts. The problem reported by linters and the difference to the golden result are also filtered and highlighted, together forming a rich context. The ADI also features a tool invocation proxy for the agents, as configuring compilation flows, simulation, and gathering results is too verbose for LLMs. Some works use LLMs to finish such tasks, but we choose to let the LLM focus on debugging.

We primarily use Verilator\cite{verilator} as the linter and simulator. However, some common error-prone patterns are legal in Verilog, which we organize the net-wires in SSA form to detect multi-driven, undriven, or partially driven wires, and unused wires and registers.

We will further explain the fault localization technique in \autoref{sec:localization} and \autoref{sec:patch} describes the mechanics of patch generation. 

\section{Patch Generation}\label{sec:patch}

\begin{figure}
    \centering
    \includegraphics[width=\columnwidth]{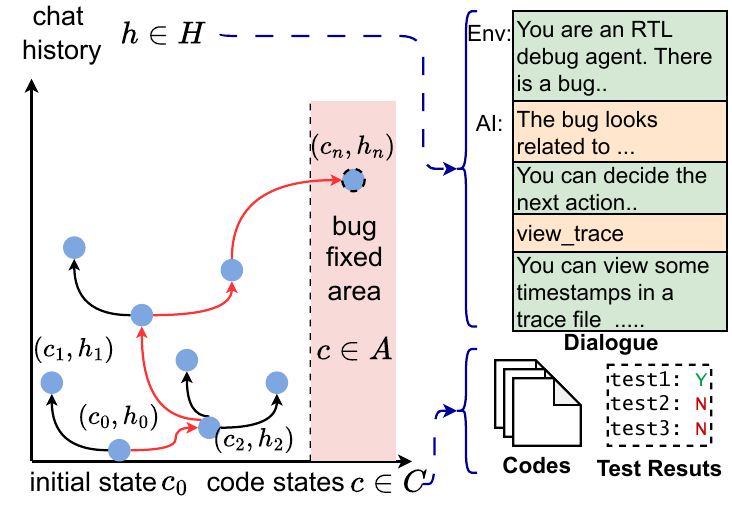}
    \caption{Search Problem Definition and the Search Tree}
    \label{fig:problem}
    \vspace{-0.5cm}
\end{figure}

\subsection{Problem Definition}

Patch generation can be defined as a search problem in the Cartesian space of the chat history $h\in H$ and code state $c\in C$. The initial state is $s_0=(c_0, h_0)$. The target is to move the code status into an area $A$ where all tests are passed. Each step will choose a state $s=(c,h)$ and run the workflow to make a new state $s'=(c', h')$. This means to reconstruct the code to $c$ and the dialog to $h$, find the position in the transition graph, and then start new rounds of conversation. The environment and agent will keep conversing for a few rounds by parsing the agent's response, taking actions, and replying with the result, until it reaches the decision node in the transition graph, where we store the new state $(c',h')$ for searching. From one state, multiple new states can be sampled, forming a tree so the agent can explore different decisions. As shown in \autoref{fig:problem}, a search tree is embedded in the $C\times H$ space. The search engine's task is to find a path on the tree from $s_0$ to a point in $A$.

\subsection{Heuristic Function}\label{subsec:value_func}

Instead of asking LLM to evaluate a state, we provide a heuristic value function for stable outcomes.
The insight is that some states in the search space are closer to the bug-fixed area.
If these points are sampled more frequently, then it is more likely to find a point in $A$. 
However, it is hard to precisely determine how far a given state is from the goal, since the location of the target region is unknown. We define a heuristic function $f(c,h)$. It estimates the likelihood that the search can succeed from a state $(c,h)$ based on prior knowledge, as shown by \autoref{equ:f}.
This function encourages passing more test benches $tb_p(c)$ and getting more information $N_Q(h)$ from the design, while avoiding compilation errors $N_{CE}(h)$ and excessive patches $N_p(h)$, token usage $N_{tok}(h)$ and getting stuck in illegal tool invocation $U(h)$. The base value $b$ adjusts the probability to balance between exploiting good nodes and exploring temporarily worse nodes. The coefficients $\lambda$ balance the strength of each factor.
The meaning of symbols is defined in \autoref{tab:symbol}. Unlike \cite{tree-of-thoughts}, which uses another LLM to evaluate the states, we find the heuristic function more controllable.

Furthermore, we choose the state to expand stochastically to balance exploration and exploitation.
In each search step, the probability of choosing a state $s=
(c,h)$ to expand is defined in \autoref{equ:softmax}. 

\begin{align}
    f(c,h) &= \lambda_1\cdot \frac{tb_p(c)}{N_{tb}}+\lambda_2\cdot N_Q(h)-\lambda_3\cdot N_{CE}(h)\nonumber\\
    &-\lambda_4\cdot N_{tok}(h)-\lambda_5\cdot U(h)-\lambda_6\cdot N_P(h)+b \label{equ:f}\\
    Pr[s&=(c,h)] = \frac{e^{f(c,h)}}{\sum_i e^{f(c_i, h_i)}}\label{equ:softmax}
\end{align}

\begin{table}[t]
\centering
\caption{Symbols in the heuristic functions}
\label{tab:symbol}
\resizebox{0.8\columnwidth}{!}{
\footnotesize
\begin{tblr}{
  row{1} = {c},
  vline{2} = {1}{},
  vline{2} = {2-15}{},
  hline{2} = {-}{},
}
Name                              & Meaning                              \\
$tb_p(c)$                & {Number of passed testbenches}      \\
$N_{tb}$                 & Number of testbenches                \\
$N_Q(h)$                   & Number of queries                    \\
$N_{CE}(h)$              & {Number of unsolved compile errors} \\
$N_{tok}(h)$             & Number of used tokens                \\
$U(h)$                      & Instability of tool invocation$^*$   \\
$N_P(h)$                   & Number of patches                    \\
$b$                         & Base value                           \\
$\lambda_i$  & Coefficients
\end{tblr}
}
\\
\footnotesize
$^*$ The frequency that the agent failed to output legal tool arguments.
\end{table}

\subsection{Stochastic Tree-of-Thoughts Algorithm}

\begin{algorithm}[t]
\caption{Search Algorithm}
\label{algo:search}
\begin{algorithmic}
\State $states \gets \left\{(c_0,h_0)\right\}$
\State $score[(c_0,h_0)] = f(c_0,h_0)$
\While{$!(timeout~||~tokens > max\_tokens)$}
   \State $(c,h)=\text{\textcolor{Blue}{sample}}(states)~\text{with}~Pr[s=(c,h)]$
   \State $score[(c,h)] = score[(c,h)] -  \text{freshness\_penalty}$
   \State $\text{checkout}(c.\text{branch})$
   \State $(c', h')=\text{\textcolor{Blue}{agent}}(c,h)$
   \State $c'.\text{branch}=\text{new\_branch}()$
   
   \If{fixed$(c',h')$}
      \State break
   \Else
      \State $state \gets state \cup \left\{(c',h')\right\}$
      \State $score[(c,h)] = f(c,h)$
   \EndIf
\EndWhile
\end{algorithmic}
\end{algorithm}

The original tree-of-thought algorithm uses DFS and BFS with beam searching, which blindly explores over the near-infinite space. Our approach uses an additional sampling process to balance between exploring deeper or wider in the tree. 

\autoref{algo:search} shows the pseudocode for the search algorithm. The algorithm maintains a set of $states$ as nodes in the search tree. The search runs repeatedly until a solution is found or the time or token budget runs out. In each step, a node is first sampled from $states$ with the distribution in \autoref{equ:softmax} (with a freshness penalty to the sampled node to encourage exploring new states). The source code is then reverted to the version associated with that state’s code branch, making that code visible to the agent. Next, the agent is invoked to run the debugging workflow from the decision node with history $h$. The agent continues until it reaches the next decision point, yielding a new state  $(c', h')$. If the new state is bug-free (passes all tests), the search terminates; otherwise, the new state is added to the set and the search continues. By carefully adjusting the coefficients in \autoref{equ:f}, the search process will expand states fairly while giving more chance for better states.

The initial states can be a set to search on a forest. For example, the set may includes a few states each with one fault candidates and one state without any to find bugs outside of the candidates. Searching on a forest is more time-consuming, so we only start with two states, based on our design scale: one with no candidates, another with the top few most possible candidates together.

\section{Fault Localization}
\label{sec:localization}

\begin{figure} [t]
    \centering
    \includegraphics[width=\columnwidth]{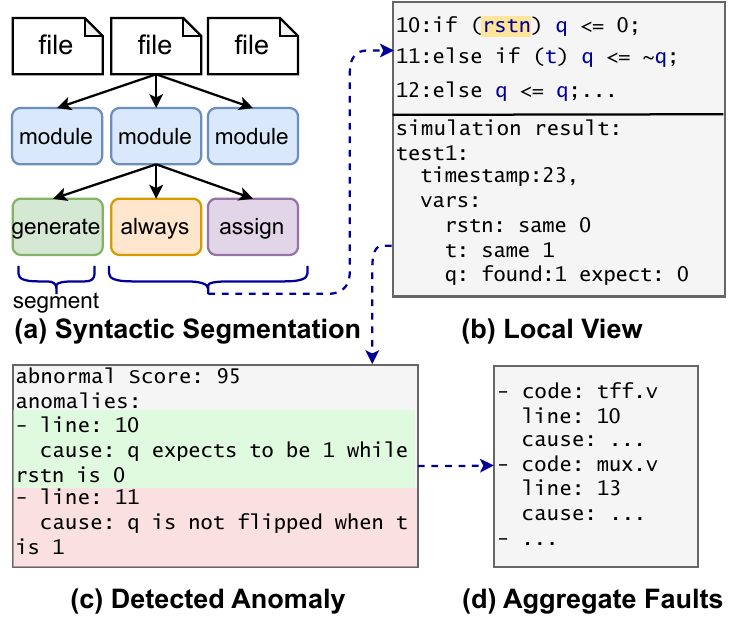}
    \caption{Detailed workflow of the Multi-Agent Anomaly Detection for Fault Localization}
    \label{fig:observer}
    \vspace{-0.5cm}
\end{figure}

The primary challenge in localizing faults within large-scale RTL designs for an LLM lies in the inherent conflict between the vast, interconnected nature of hardware code and the LLM's finite input context window, not to mention reserving enough space for outputs. The multi-agent anomaly detection technique is proposed to overcome this. The core principle is to decompose the global fault localization problem into a series of localized, parallelizable anomaly detection tasks. Each task is assigned to an independent LLM-based agent that analyzes a small, contextually relevant fragment of the design, thereby circumventing the context limitations and enabling scalable analysis.

This framework is shown in \autoref{fig:observer}. First(a), the complete RTL design is partitioned into multiple smaller, syntactically complete code segments by recursively partitioning on the AST until some sibling nodes are small enough.
Second(b), they are paired with the relevant error message, including lintings and waveform comparison results, forming a self-contained "local view". Third(c), each local view is independently processed by an LLM agent. The agent is prompted to assess the likelihood that the code segment is related to the error. Fourth(d), the outputs from all agents are aggregated. Code segments with a score higher than a predefined threshold are classified as anomalies, which are then sorted for the top-k and filtered again, yielding a refined set of high-confidence locations.

For instance, in the example depicted in \autoref{fig:observer}, a code segment from a multiplexer is presented with a simulation mismatch for the signal $q$. An LLM agent identifies two suspicious lines (Line 10 and 11), assigning a high score, indicating this segment is an anomaly. After aggregating and sorting, three anomalies were left, including this one. The final filtering stage correctly discards Line 11 as illogical, pinpointing Line 10 as the true fault location. This is because the real cause of the bug is that, the $rstn$ signal should be functional when it is 0 instead of 1 at line 10, and consequently it affects the "else" statements at line 11.

\section{Experiments} \label{sec:experiment}

\subsection{Experiment Setup}

We evaluate using the RTL-repair\cite{rtl-repair} dataset, which contains a variety of buggy RTL designs, widely used in prior works\cite{rtl-repair,cirfix}. The roots of bugs varied from structural to algorithmic. Our results are compared against four baselines. For traditional APR, \textit{RTL-repair}\cite{rtl-repair} is the state-of-the-art, which transforms program repair into SMT problems. For software LLM-based APR, \textit{SWE-Agent}\cite{swe-agent} is the state-of-the-art with general ability to communicate with the environment. For LLM-based APR designed for RTL, MEIC\cite{MEIC} and UVLLM\cite{uvllm} are recent works that are open-sourced. In our evaluation, a case is passed if the waveform is the same as the golden reference, the same as prior works.

We evaluate our approach with Deepseek-V3\cite{deepseekv3} API calls. Only the source code is modifiable to avoid the agent removing the tests. We use Verilator\cite{verilator} to lint and simulate the designs, and build additional linting on top of Yosys \cite{yosys} processed net list.
To estimate the reliability, we use the pass@k metric\cite{evaluating}, which means the probability that the bug is fixed within k independent retries within the given time and resource limit. 
We sample pass@k with 10 retries, each with a 10-minute timeout and 1M total token limit. Because token usage can only be checked after the LLM finishes inference, the actual time and token usage may be slightly more than the limit. The baselines do not support such a limit, so we adjust the number of rounds and retries for a similar time usage.

\subsection{Bug Repair Results}

\begin{table*}
\centering
\caption{Evaluation on Bug Repairing Ability for RTL-repair\cite{rtl-repair}, SWE-Agent\cite{swe-agent}, MEIC\cite{MEIC}, UVLLM\cite{uvllm} and R$^3$A. \\The first 17 benchmarks are easy cases with only one module and one file, and the rest are much more complex. \\ Symbols: \ding{51} test passed. \ding{53} test failed. \Circle no result(not applicable or timed out). P: total passed tests}
\label{tab:fix}
\resizebox{\linewidth}{!}{
\begin{tblr}{
  vline{2,19,34,38} = {-}{},
  hline{2,3,4,5,6} = {-}{},
}
           & 1 & 2 & 3 & 4 & 5 & 6 & 7 & 8 & 9 & 10 & 11 & 12 & 13 & 14 & 15 & 16 & 17 & 18 & 19 & 20 & 21 & 22 & 23 & 24 & 25 & 26 & 27 & 28 & 29 & 30 & 31 & 32 & P \\
\cite{rtl-repair} & \ding{51} & \ding{53} & \ding{51} & \Circle   & \ding{51} & \ding{51} & \ding{51} & \Circle   & \ding{51} & \ding{51} & \ding{51} & \ding{51} & \ding{51} & \ding{53} & \Circle   & \ding{51} & \ding{51} & \Circle   & \Circle   & \Circle   & \ding{51} & \Circle   & \Circle   & \Circle   & \Circle   & \Circle   & \Circle   & \ding{51} & \ding{51} & \Circle   & \Circle   & \ding{51} & 16     \\
\cite{swe-agent}  & \ding{53} & \ding{53} & \ding{51} & \ding{51} & \ding{51} & \ding{51} & \ding{53} & \ding{51} & \ding{51} & \ding{51} & \ding{53} & \ding{53} & \ding{51} & \ding{51} & \ding{51} & \ding{51} & \ding{51} & \ding{53} & \ding{53} & \ding{53} & \ding{53} & \ding{53} & \ding{53} & \ding{53} & \ding{53} & \ding{51} & \ding{53} & \ding{51} & \ding{51} & \ding{53} & \ding{53} & \ding{53} & 15 \\
\cite{MEIC}       & \ding{51} & \ding{51} & \ding{53} & \ding{53} & \ding{53} & \ding{51} & \ding{51} & \ding{51} & \ding{51} & \ding{51} & \ding{51} & \ding{53} & \ding{53} & \ding{53} & \ding{51} & \ding{53} & \ding{53} & \ding{53} & \ding{53} & \ding{53} & \ding{53} &  \ding{53} & \ding{53} & \ding{53} & \ding{53} & \ding{51} & \ding{53} & \ding{51} & \ding{51} & \ding{53} & \ding{53} & \ding{53} & 12 \\
\cite{uvllm}      & \ding{51} & \ding{51} & \ding{53} & \ding{53} & \ding{53} & \ding{51} & \ding{51} & \ding{51} & \ding{51} & \ding{51} & \ding{51} & \ding{51} & \ding{51} & \ding{51} & \ding{51} & \ding{51} & \ding{51} & \ding{53} & \ding{53} & \ding{51} & \ding{51} & \ding{53} & \ding{53} & \ding{53} & \ding{53} & \ding{51} & \ding{51} & \ding{51} & \ding{51} & \ding{53} & \ding{53} & \ding{53} & 20    \\
R$^3$A            & \ding{51} & \ding{51} & \ding{51} & \ding{51} & \ding{51} & \ding{51} & \ding{51} & \ding{51} & \ding{51} & \ding{51} & \ding{51} & \ding{51} & \ding{51} & \ding{51} & \ding{51} & \ding{51} & \ding{51} & \ding{51} & \ding{51} & \ding{51} & \ding{51} & \ding{53} & \ding{51} & \ding{51} & \ding{51} & \ding{51} & \ding{51} & \ding{51} & \ding{51} & \ding{53} & \ding{51} & \ding{53} & 29 
\end{tblr}
}
\footnotesize
Benchmarks:
1. decoder\_w1 2. decoder\_w2 3. counter\_k1 4. counter\_w1 5.counter\_w2 6. flop\_w1 7. flop\_w2 8. fsm\_w1 9. fsm\_w2

10.fsm\_s1 11.fsm\_s2 12.shift\_w1 13.shift\_w2 14.shift\_k1 15.mux\_k1 16.mux\_w1 17.mux\_w2 18.sha3\_r1 19.sha3\_w1 20.sha3\_w2 21.sha3\_s1

22.pairing\_w1 23.pairing\_w2 24.pairing\_k1 25.reed\_b1 26.reed\_o1 27.sdram\_w1 28.sdram\_w2 29.sdram\_k1 30.i2c\_w1 31.i2c\_w2 32.i2c\_k1
\end{table*}

We first evaluate the ability to reliably fix a bug within the retry, time, and resource limit. As shown in \autoref{tab:fix}, our approach can fix 29 of the bugs, while \textit{RTL-repair}, \textit{SWE-Agent}, and \textit{UVLLM} can only fix 16, 15, 12, and 20 of them, respectively. 

The first 17 benchmarks are easy cases with only one module and one file. The LLM understands them very well and can fix them easily. However, some of the designs have some differences from commonly used textbook designs, e.g., the 3-to-8 decoder has an enable signal that is not useful. This causes the agent to be confused in the first few rounds, but it can be recovered after validating the agent's understanding by trial-and-error. 

The latter 15 benchmarks are much more complex. Their root cause is hidden deep inside the module tree and file structure. 
Some faults can be found by linters. \textbf{sha3\_r1}, \textbf{pairing\_k1}, and \textbf{reed\_b1} have incorrect width on signals or expressions, which can trigger warnings in Verilator. \textbf{sha3\_w1} has a wire partially undriven, and \textbf{pairing\_w2} mistakenly exchanges the input and output of a module. Though legal in Verilog, these error-prone cases can be detected by our additional linter. 
Others can benefit from the fault localization technique: \textbf{sha3\_w2} miswrites a condition into 0, and \textbf{pairing\_w1} mistakenly updates a loop variable, causing a dead loop.
Our fault detection fails to locate the fault of \textbf{sha3\_s1}, which drops an update condition when the buffer is full. However, from the waveform, the patch generation agent can still detect and fix the error that happens exactly when the full signal changes.

We fail to fix the \textbf{pairing\_w1},  \textbf{i2c\_w1} and \textbf{i2c\_k1} case. The dead loop in \textbf{pairing\_w1} is an algorithmic error that is hard to solve without knowing the details of the Tate pairing algorithm. \textbf{i2c} benchmarks require simulating in-cycle latency, which is a different category of problem.

For \textit{RTL-repair}, many bugs fall out of its predefined templates. For example, the bug of \textbf{decoder\_w2}, \textbf{fsm\_w1}, \textbf{counter\_w1}, and \textbf{shift\_k1} benchmarks has some wrong "if-else" or "case" structures or wrong sensitivity lists in an "always" block that its fix range simply cannot cover them.
\textit{SWE-Agent} is not adapted to RTL and therefore wastes too much time on tooling. \textit{MEIC} ranks RTL codes with a scorer agent, which further increases the instability.  \textit{UVLLM} localizes fault by Verilator linting and Strider signal tracing, and samples multiple LLM outputs from the same input. This strategy works well in smaller cases, but only linting works in more complex multi-file cases. Besides, \textit{UVLLM} requires a problem description text as input, but our setting assumes no specification. If specification is provided, then it can pass the simple cases of \textbf{counter}, but has no improvement in other cases. For the \textbf{counter} cases, signal tracing fails to trace through combinational logics, and patch by experience also fails because the design is uncommon: the "overflow" signal is not related to the "enable" signal. Without specification, the LLM is hallucinated to overwrite it.

We found that the original testbench in RTL-repair for \textbf{sdram\_w2} does not expose the error, and the "bug" of \textbf{sdram\_k1}, \textbf{reed\_o1} and \textbf{fsm\_s1}, and \textbf{fsm\_w2} has no effect. We keep them unchanged for a fair comparison with RTL-repair.

In conclusion, our approach can reliably fix the bugs within the time limit, passing 45\% and 81.3\% more bugs than LLM-based\cite{swe-agent, MEIC, uvllm} and traditional methods\cite{rtl-repair}, respectively. \textit{RTL-repair} has limited fix capability, \textit{SWE-Agent} is not adapted to RTL, while \textit{MEIC} and \textit{UVLLM} do not address the characteristics of LLM for a reliable outcome.

\subsection{Ablation Study of Reliability}\label{subsec:reliable}

\begin{figure} [t]
    \centering
    \includegraphics[width=\columnwidth]{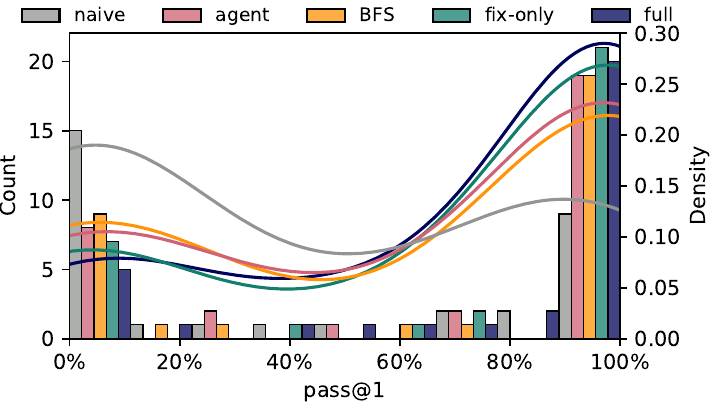}
    \caption{Comparison of pass@1 rate between different versions. The figure is a combined figure of a histogram with Kernel Density Estimation (KDE) lines. If the line is higher on the right side, the method is generally more reliable.}
    \label{fig:reliable}
    \vspace{-0.5cm}
\end{figure}

To further evaluate the reliability of our approach, we perform an ablation study with five settings. The \textit{naive} setting simply requests the LLM to fix the bug given observed faults. The \textit{agent} setting implements the repair agent without tree-of-thoughts. The \textit{BFS} setting implements the repair agent with tree-of-thoughts and its original BFS algorithm\cite{tree-of-thoughts}  (with our heuristic). DFS is not applicable as there is no reasonable stop criterion other than fixing the bug. The \textit{fix-only} setting implements our search algorithm, without fault localization. The \textit{full} setting uses all proposed techniques.

As shown in \autoref{fig:reliable}, we evaluate the pass@1 metric of these methods. The higher the pass@1 is, the more reliable the method is. The histogram shows the number of benchmarks with a pass@1 that fall within a range. The 0\% bin (the bug can not be fixed) shows that the \textit{full} setting fixes most bugs, followed by \textit{fix-only}, \textit{agent}, \textit{BFS} and \textit{naive}. The kernel density estimation line shows the smoothed distribution of benchmarks with different pass@1. The simple cases are almost 100\% solvable and do not make much difference. The number may jitter due to limited samples. The improvement of reliability mainly increases the pass@1 of difficult cases from negligible to non-negligible. Overall, \textit{full} setting is the most reliable one as the right side of the KDE line is the highest, then \textit{fix-only} is less reliable,  followed by \textit{agent} and \textit{BFS}, and \textit{naive}.

\begin{figure} [t]
    \centering
    \includegraphics[width=\columnwidth]{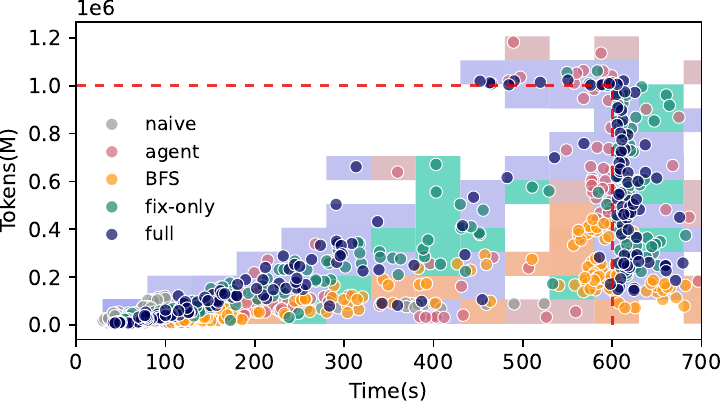}
    \caption{Comparison of the time and token usage between different versions. Each point is one test. The unit for tokens is millions, and the unit for time is seconds. The red line marks the time and token budget range, where a point fails if falling out of range. }
    \label{fig:timetoken}
    \vspace{-0.5cm}
\end{figure}

\autoref{fig:timetoken} shows a comparison of time and token usage to analyse the reason for the difference in reliability. The usage of \textit{naive} mostly comes from fault localization. The \textit{BFS} algorithm uses too few tokens before time runs out, as its time is spent on nodes closer to the root, which explores too much but does not exploit the good nodes enough. The naive \textit{Agent} algorithm, on the contrary, always expands the furthest node, consuming more tokens within the time limit, because it has no exploration ability and sometimes runs into dead ends. The \textit{full} setting balances exploration and exploitation with stochastic state expansion. The \textit{fix-only} setting uses more time (213.27s) than \textit{full} (258.64s) on average. This is because the fault localization result saves rounds for the repair agent with global views.

In conclusion, comparison with \textit{naive}, \textit{BFS} and \textit{agent} demonstrates that the stochastic tree-of-thought approach greatly improves reliability, and is more likely to search for a valid solution. Comparison with \textit{fix-only} shows that the multi-agent fault localization approach improves overall reliability. However, in some benchmarks such as \textbf{counter\_w1} and \textbf{sha3\_w2}, it slightly reduces the pass@1 rate. This is because observing local views is sometimes noisy due to insufficient context. The overall average pass@1 and pass@5 rate for the \textit{full} setting are 75.0\% and 86.7\%, respectively.

\section{Conclusion}

Traditional APR cannot cover all the bugs in RTL, while directly applying LLM meets challenges of stochasticity and input quality. We propose \rdba, a reliable RTL repairing framework, including the multi-agent anomaly detection method for fault localization, 
and the stochastic tree-of-thought repair agent for patch generation, 
turning the LLM’s randomness into reliable search and consistently producing valid fixes. 
Experiments show that our approach can fix 90.6\% of bugs and achieves an 86.7\% pass@5 rate on average.

\newpage

\bibliographystyle{ACM-Reference-Format}
\bibliography{bib.bib}

\end{document}